# Parallel K-Medoids++ Spatial Clustering Algorithm Based on MapReduce

Xia Yue

*School of Geosciences and Info-Physic, Central South University, Changsha 410083, China*
*xiayuecsu@163.com*

Wang Man*

*Department of Spatial Information Science and Engineering, Xiamen University of Technology, Xiamen 361024, China*
*mwang.xmut@gmail.com*
**Corresponding Author*

Jun Yue

*Institute of Remote Sensing and Geographic Information System, Peking University, Beijing 100871, China,*
*jyue@pku.edu.cn*

Guangcao Liu

*Xiamen Great Power GEO Information Technology Co., Ltd., Xiamen 361008, China*
*gcliu.xmgeo@gmail.com*

Clustering analysis has received considerable attention in spatial data mining for several years. With the rapid development of the geospatial information technologies, the size of spatial information data is growing exponentially which makes clustering massive spatial data a challenging task. In order to improve the efficiency of spatial clustering for large scale data, many researchers proposed several efficient clustering algorithms in parallel. In this paper, a new K-Medoids++ spatial clustering algorithm based on MapReduce for clustering massive spatial data is proposed. The initialization algorithm to decrease the number of iterations is combined with the MapReduce framework. Comparative Experiments conducted over different dataset and different number of nodes indicate that the proposed K-Medoids spatial clustering algorithm provides better efficiency than traditional K-Medoids and scales well while processing massive spatial data on commodity hardware.

## 1. Introduction

Clustering is one of the common problems in data mining and computational geometry. It enables to group a set of multidimensional points into different classes so that different multidimensional points with inter-point similarity and intra-point difference can be classify into the same cluster [1]. In other words, clustering is to find natural clusters in the data set which has been widely used in the areas such as city planning, remote sensing and geographic information science (GIScience) [2]. Spatial clustering in this work is to classify the two dimensional spatial points in the area of GIScience.

However, traditional K-Means clustering algorithm proposed by Refs. 3-4 in different research areas independently is sensitive to outers. As a result, the K-Medoids clustering algorithm is proposed which is more robust than K-Means, where the medoids are used to replace the mean of the multidimensional points in the same class. The replacement rule of the medoid of a specific cluster is based on the cost and the non-medoid with the less cost will replace the current medoid which makes K-Medoid less sensitive to noises compared with the K-Means clustering [5].

Spatial clustering algorithms is efficient while classifying the small size spatial data set but the efficiency will drop sharply as the data volume increasing. As for K-Medoids clustering algorithm, the performance is good when processing small size data set but the performance will decrease when the scale of the spatial data is increased. However, the size of the spatial data processed by many spatial analysis applications will frequently surpass the peta-scale threshold with the rapid development of geospatial information technique and the computational requirements will increase enormously [6]. Improving the spatial clustering algorithm itself and optimizing the proceeding of execution are commonly believed to be a valid solution to increase the efficiency while clustering massive spatial data. The K-Medoids clustering algorithm is time consuming while processing massive two dimensional spatial points though it is robust against outers. As a result, the more efficient K-Medoids spatial clustering algorithm should be proposed and this paper pays attention on this issue.

MapReduce is a parallel programming architecture for processing massive dataset. Google and Apache Hadoop both provide MapReduce framework with dynamic support and fault tolerance [7]. In this work, the K-Medoids spatial clustering algorithm is adapted to MapReduce framework so that the job of clustering massive spatial data can be implemented on Hadoop platform. What's more, the initialization algorithm proposed in Ref. 8 to decrease the number of iterations is combined with the MapReduce framework and the proposed algorithm is termed K-Mediods++. By designing the proper

Map and Reduce procedure, the proposed K-Medoids++ can be executed effectively on multi-node cluster. The experiments on Hadoop and HBase conducted over different dataset and different number of nodes demonstrate that the proposed K-Medoids spatial clustering algorithm can provide better efficiency while processing massive spatial data on commodity hardware than traditional K-Medoids and scales well.

The remaining parts of this paper are organized as follows. In Section 2, MapReduce, Hadoop, HBase and the traditional K-Medoids algorithm are briefly introduced. Section 3 presents the basic MapReduce process and the proposed K-Medoids spatial clustering algorithm. Section 4 shows the experiments with respect to efficiency and speedup. In Section 5, some conclusions are given.

## 2. Preliminaries

In this section, the core concepts about MapReduce, Hadoop, HBase and K-Mediods algorithm is given.

### 2.1. *MapReduce*

MapReduce is a distributed programming model and an efficient scheduling model for task resource allocation originally from functional programming language. The main ideology of this model is using map and reduce functions to spilt and combine the data. A MapReduce program is composed of a Map function which is designed for sorting and filtering the input data and a Reduce function which is designed for summing up the data. The MapReduce framework is designed for managing the distributed servers and running massive computing jobs in parallel. It manages all data transfers between different tasks of the framework with redundancy and high fault tolerance. The basic programming model of MapReduce architecture is shown in Fig. 1.

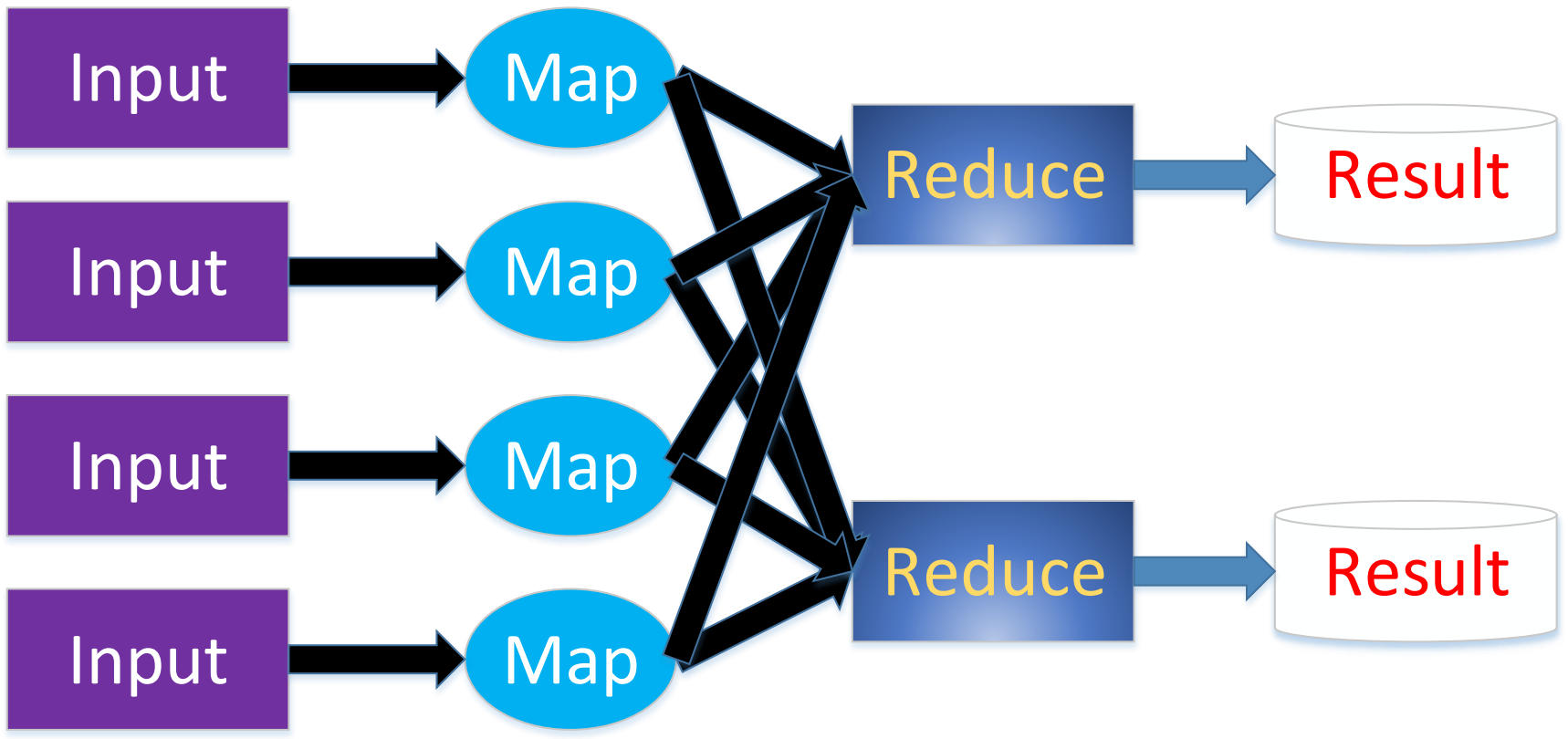

Fig. 1. Basic programming model of MapReduce

MapReduce frameworks have the ability of processing large distributed tasks in parallel environment such as cloud computing infrastructures. It supports the scientific researches as well as the commercial applications by building large clusters to support data-intensive calculations. With the introducing of MapReduce programming model, designing the map function which processes a <key, value> pair to generate a <key, value> pair and the reduce function which takes all combined <key, value> pairs together to generate a set of outputs is of vital importance to process the massive data [9].

### 2.2. *Hadoop and HBase*

Hadoop is a Java-based implementation of MapReduce programming framework by Apache for massive data distributed processing on cloud computing clusters. The basic feature of Hadoop is that it can automatically handle the hardware failure so that the cloud computing clusters can be built on commodity hardware. Hadoop consists of operation level component, Hadoop Distributed File System (HDFS) and MapReduce component. HDFS is used for data storage and distributing the data into different blocks in different clusters. HDFS ensures the high fault tolerance by replicating the data blocks into different nodes and the data locality information in HDFS is used for the communication between the nodes.

A typical instance of Hadoop includes a master node and several slave nodes. The master node consists of a JobTracker, TaskTracker, NameNode, and DataNode. The JobTracker which is the service for assigning the jobs to the nodes with the target data and the TaskTracker which is a node in the cluster that accepts jobs from a JobTracker are the components of the MapReduce Engine. A slave node mostly acts as a TaskTracker and DataNode. NameNode is the control node and DataNode is the work node of HDFS. The structure of Hadoop cluster is shown in Fig. 2.

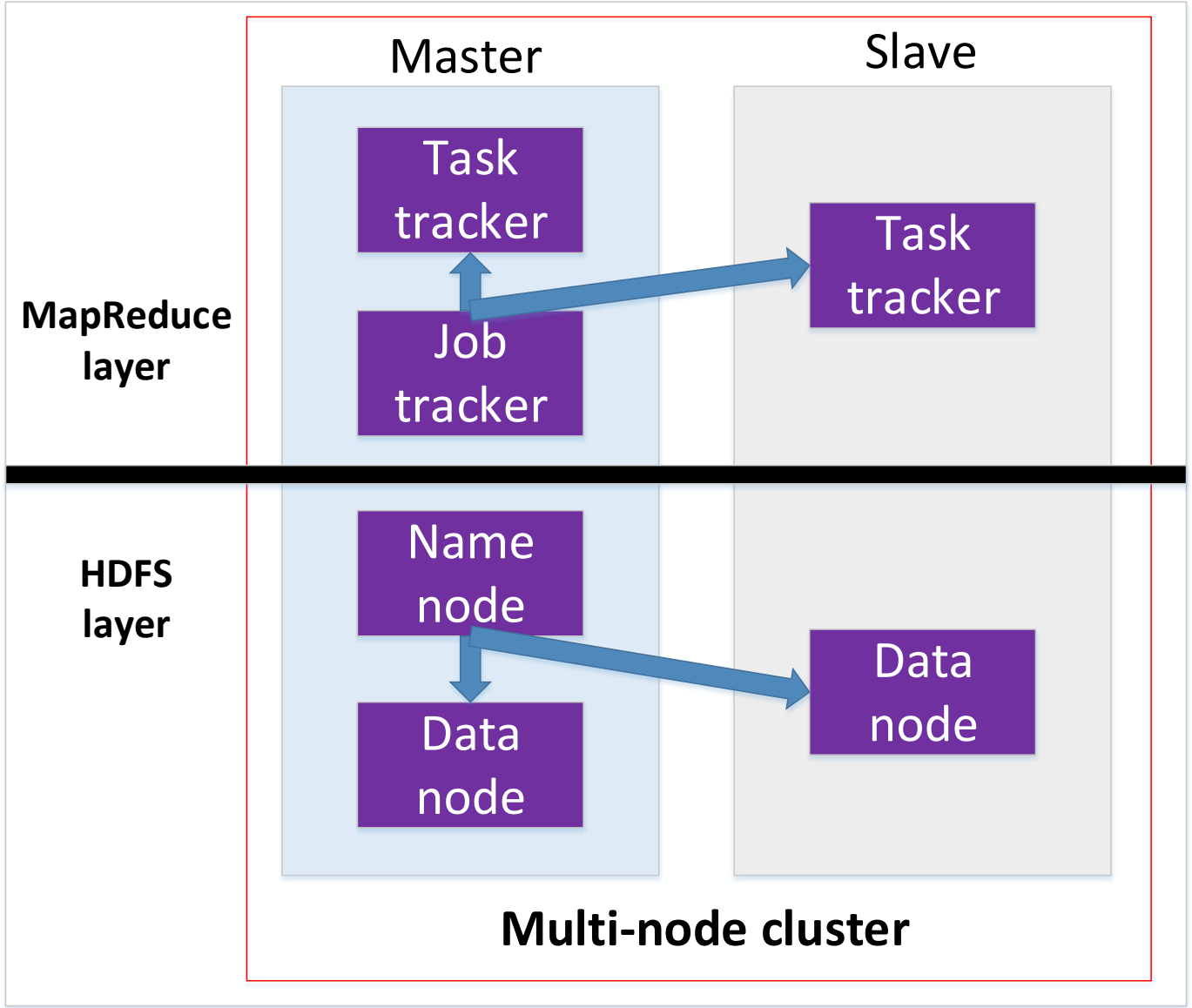

Fig. 2. Multi-node cluster

HBase is an additional software package of Hadoop and is built on top of HDFS. It is a part of Apache Hadoop project providing the functions like BigTable with the feature of fault tolerance. The data is stored in a number of rows with unique keys in HBase but each row can store any number of columns. For example, one row can have one column while another row can have thousands of columns. The data in HBase is stored in distributed nodes in the cluster. The basic storage unit is HStore with each family stored in it. For example, column A is stored in the first HStore, column B is stored in the second HStore, column C and column D are stored in the last HStore. It shows the efficiency of retrieving a column rather than a row in HBase. HMaster is used to request the HStore to retrieve a column [10].

## 2.3. *K-Medoids*

The K-Means clustering has become one of the popular clustering algorithm because of its excellent performance. However, it is easily influenced by the outliers. In order to improve the robustness of the clustering algorithm, the K-Medoid algorithm is proposed by using the medoids which are the elements in the date set as the center of the clusters. It attempts to minimize the cost between non-medoids of the cluster and the medoid of the cluster. PAM (Partitioning Around Medoids), the earliest K-Medoids algorithm, was proposed by Kaufman and Rousseeuw [11] which can be described as follow. (1) Select $k$ points in the data set arbitrarily as the initial medoids. (2) Classify each non-medoid point to the nearest medoid according to the Euclidean distance. (3) Choose the new medoid with the least cost for each cluster. (4) Continue Step (2) until the total cost remain the same. The total cost can be formulated as follows.

$$E = \sum_{n=1}^{k} \sum_{p_m \in C_n} |p_m - o_n|^2 \quad (1)$$

where $C_n$ and $o_n$ are the $n$th cluster and the medoid of this cluster, separately. $p_m$ is the point in the $n$th cluster. $|p_m - o_n|^2$ is the Euclidean distance between one of the non-medoid and the medoid in the same cluster. There are for cases for the replacement between the non-medoid $O_{current}$ and the medoid $O_i$ and $O_j$. Case 1: p is classified to medoid $O_i$. If the closest medoid is another $O_j(i \neq j)$ after $O_{current}$ replaced $O_i$, then classify p to $O_j$. Case 2: p is classified to medoid $O_i$. If the closest medoid is $O_{current}$ after $O_{current}$ replaced $O_i$, then classify p to $O_{current}$. Case 3: p is attached to medoid $O_j$. If the closest medoid is $O_j$ after $O_{current}$ replaced $O_i$, then the label of p remain the same. Case 4: p is attached to medoid $O_j$. If the closest medoid is $O_{current}$ after $O_{current}$ replaced $O_i$, then classify p to $O_{current}$ [12].

The K-Medoid algorithm clustering algorithm works effectively while clustering low dimensional data sets with small size, but is not that efficient for massive data sets because of its complexity [13]. As a result, the paralleled K-Medoids++ algorithm is proposed with the effective initialize algorithm.

# 3. Methodology

## 3.1. *Initialization algorithm of K-Medoids++*

The K-Medoids algorithm finds cluster medoids by minimizing the intra-class discrepancy. However, the k-Mediods algorithm has the following theoretical drawbacks.

First, the number of medoids should be given in advance. However, the number of medoids is hard to determine in many cases. Second, the initial medoids should be selected manually which will increase the number of iterations.

The proposed k-Mediods++ algorithm addresses the second concern by using the initialization algorithm proposed in Ref. 8 to decrease the number of iterations. The explicit algorithm is as follows.

(1) The initial medoid is chosen randomly among all of the spatial points.

(2) For each spatial point $p$, compute the distance between $p$ and the nearest medoids which is termed as $\mathrm{D}(p)$ and sum all the distances to $S$ .

(3) The next medoid is determined by using weighted probability distribution. Specifically, a random number $R$ between zero and the summed distance $S$ is chosen and the corresponding spatial point is the next medoid.

(4) Step (2) and Step (3) are repeated until $k$ medoids have been chosen.

### 3.2. *Basic MapReduce data processing Flow*

The basic MapReduce algorithm for K-Medoids++ spatial clustering can be summarized to three steps as follows [13, 14].

(1) Generating $k$ points as initial $k$ medoids using the initialization algorithm.

(2) The Map procedure on Hadoop.

Map: (key, value) -> list (key, value).

The keys of the output are identifiers of clusters and the values of the output are the coordinates of spatial points.

(3) The Reduce procedure on Hadoop.

Reduce: (key, list(value)) -> list (key, value)

The keys of the input and the output are identifiers of clusters, and the values of the input are the coordinates of spatial points. The values of the output are coordinates of the new medoids.

The three steps can be simplified as follows.

(1) The initial medoids are generated.

(2) Applying K-Medoids in MapReduce architecture.

### 3.3. *Parallel algorithm implement*

The implement of the parallel algorithm needs the MapReduce job. The map function acts as the operation of assigning each spatial points to the closest medoid while the reduce function acts as the operation of generating the new medoid.

The input spatial data is stored on HBase as a sequence file of coordinates. The key of map function is the row number in the HBase dataset and the value is a string of the corresponding coordinate. The spatial points are split and sent to mappers. As a result, the distance between the medoid and the non-medoid is computed in parallel. All the initial medoids are stored in the file of medoids as the input of the map function. The output is the (key, value) pair containing the information of the cluster ID and the coordinate. The pseudocode of Map function is shown in Table 1.

Table 1. Pseudocode of Map function

| Algorithm 1. Map(ImmutableByteWritable row, Result value, Context CText) |
|---|
| Input: The row in HBase and the file of medoids. |
| Output: < Out-Key , Out-Value> pair. Out-Key is the cluster ID. Out-Value is the coordinate of this row. |
| 1. Get the coordinate from values.<br>Point point=GetCoordinateFromText(Pointlist.next()); |
| 2. Initialize the minimum of the distance as the max value of double type.<br>MinOfDistance = Double.MAX VALUE; |
| 3. Classify the available points |
| Medoids=LoadMedoids(FILE);<br>for(int i = 0;i<NumofMedoids(Medoids); i++) {<br>Medoid=GetMedoidFromClusterID(Medoids,i);<br>Distance = CalculateDistance(Medoid,point);<br>If(Distance < MinOfDistance)<br>{ MinOfDistance = Distance; Out-Key =(Text)i; }<br>} |
| 4. Output the cluster ID and the coordinate of the point.<br>CText.write(Out-Key, values); |
| 5. Construct value' as a string comprise of the values of different dimensions; |
| 6. output < key, value> pair; |
| 7. End |

The input of the reduce function is the output list of (key, value) pairs of the map function. The (key, value) pairs with the same key value are combined in the list. The costs of the medoid and the candidate medoids are calculated and the candidate medoids with the least cost is chosen as the new medoid for the next iteration. The output is the (key, value) pair containing the information of the cluster ID and the coordinate of the new medoid. The pseudocode of Reduce function is shown in Table 2.

Table 2. Pseudocode of Reduce function

| Algorithm 2. Reduce (Text Key, Iterable<Text> Pointlist, Context CText) |
|---|
| Input: Key is the cluster ID. Pointlist is the list of the coordinates assigned to this cluster. CText is for transferring the data. |
| Output: < Out-Key , Out-Value> pair. Out-Key is the same cluster ID with Key. Out-Value is the new medoid. |
| 1. Get the cluster ID from the input key.<br>Cluster_ID= GetClusterID(Key); |
| 2. Load the file of the mediods and calculate the cost of the medoid of the specific cluster.<br>Medoids=LoadMedoids(FILE);<br>Medoid=GetMedoidFromClusterID(Medoids,Cluster_ID);<br>MinCost= CalculateCost(Mediod); |
| 3. Calculate the cost of the candidate medoids and choose the corresponding cluster ID.<br>i=0;<br>while(Pointlist.hasNext()){<br>Point point=GetCoordinateFromText(Pointlist.next());<br>Cost=CalculateCost(point); i++<br>If(MinCost < Cost)<br>{ MinCost = Cost; Out-Value=(Text)point;}} |
| 4. Output the cluster ID and the coordinate of the new medoid.<br>CText.write((Text)Cluster_ID, Out-Value); |
| 5. Update the file of the mediods.<br>writefile((Text)Cluster_ID, Out-Value); |

(3) Compare the previous file of the mediods with the updating file of the mediods. If the medoids retain the same, then the program output the clustering result. Otherwise, go back to another iteration with the new file of medoids.

## 4. Experiment

### 4.1. *Experimental environment*

Seven VMware virtual nodes with Ubuntu (32-Bit) 10.04 operating system running on three host computers are used in this experiment and the detailed configurations of the three host computers are shown in Table 3. The cloud computing cluster based on Hadoop consists of six slave nodes and one master nodes. The six slave nodes are named slave01 to slave06 specifically and each slave node acts as DataNode, TaskTracker and HRegionServer while the master node acts as NameNode, JobTracker and HMaster.

Table 3. Configurations of nodes

| Node | CPU | CP[a] | RAM(GB) | Host |
|---|---|---|---|---|
| Master | Intel i5-3210M | 4 | 8 | Host1 |
| Slave01<br>Slave02<br>Slave03 | AMD A8-5600K | 2 | 8 | Host2 |
| Slave04<br>Slave05<br>Slave06 | Intel E7500 | 2 | 2 | Host3 |

[a]CP=number of cores per processor

### 4.2. *Experimental result*

In this section, the performance of the proposed spatial clustering algorithm is evaluated in terms of speedup and efficiency on the cluster mentioned above. The speedup is to evaluate the expansibility of the proposed algorithm which indicates the ability of processing the same size of dataset as the hardware resources increasing. To evaluate the speed up, the dataset remained the same and the number of nodes increased. The linear speedup is the ideal speed up of the parallel algorithm. For example, a parallel system with n times the number of nodes yields a speedup of n. However, linear speedup is hard to realize because the cost of communication between computer nodes is increasing as the number of nodes growing.

Six groups of cluster is designed to evaluate the performance of the proposed algorithm and each group of cluster has certain number of cloud computing node. The specific composition of each group of cluster is shown in Table 4. Three spatial data sets with data sizes and numbers of spatial points are shown in Table 5.

Table 4. Cluster Composition

| Cluster | Members |
| --- | --- |
| 7 Nodes | Master, Slave01, Slave02, Slave03, Slave04, Slave05, Slave06 |
| 6 Nodes | Master, Slave01, Slave02, Slave03, Slave04, Slave05 |
| 5 Nodes | Master, Slave01, Slave02, Slave03, Slave04 |
| 4 Nodes | Master, Slave01, Slave02, Slave03 |

Table 5. Datasets with different sizes

| Dataset | Data Size | Number of spatial points |
| --- | --- | --- |
| Dataset 1 | 515MB | 1316792 |
| Dataset 2 | 958MB | 2449101 |
| Dataset 3 | 1259MB | 3220460 |

The experiments are conducted on datasets with different sizes. The number of cloud computing nodes varied from 4 to 7. The sizes of three spatial data sets are 515MB, 958MB and 1259MB specifically. The execution time of the proposed K-Medoids spatial clustering algorithm is given in Table 6. A visual illustration with a histogram is shown in Fig. 3.

Table 6. Time consuming of proposed algorithm over different sizes of datasets

| Cluster | Dataset 1 | Dataset 2 | Dataset 3 |
| --- | --- | --- | --- |
| 4 Nodes | 532072ms | 891090ms | 1037331ms |
| 5 Nodes | 464354ms | 784585ms | 860312ms |
| 6 Nodes | 418680ms | 721358ms | 785269ms |
| 7 Nodes | 399054ms | 700821ms | 747987ms |

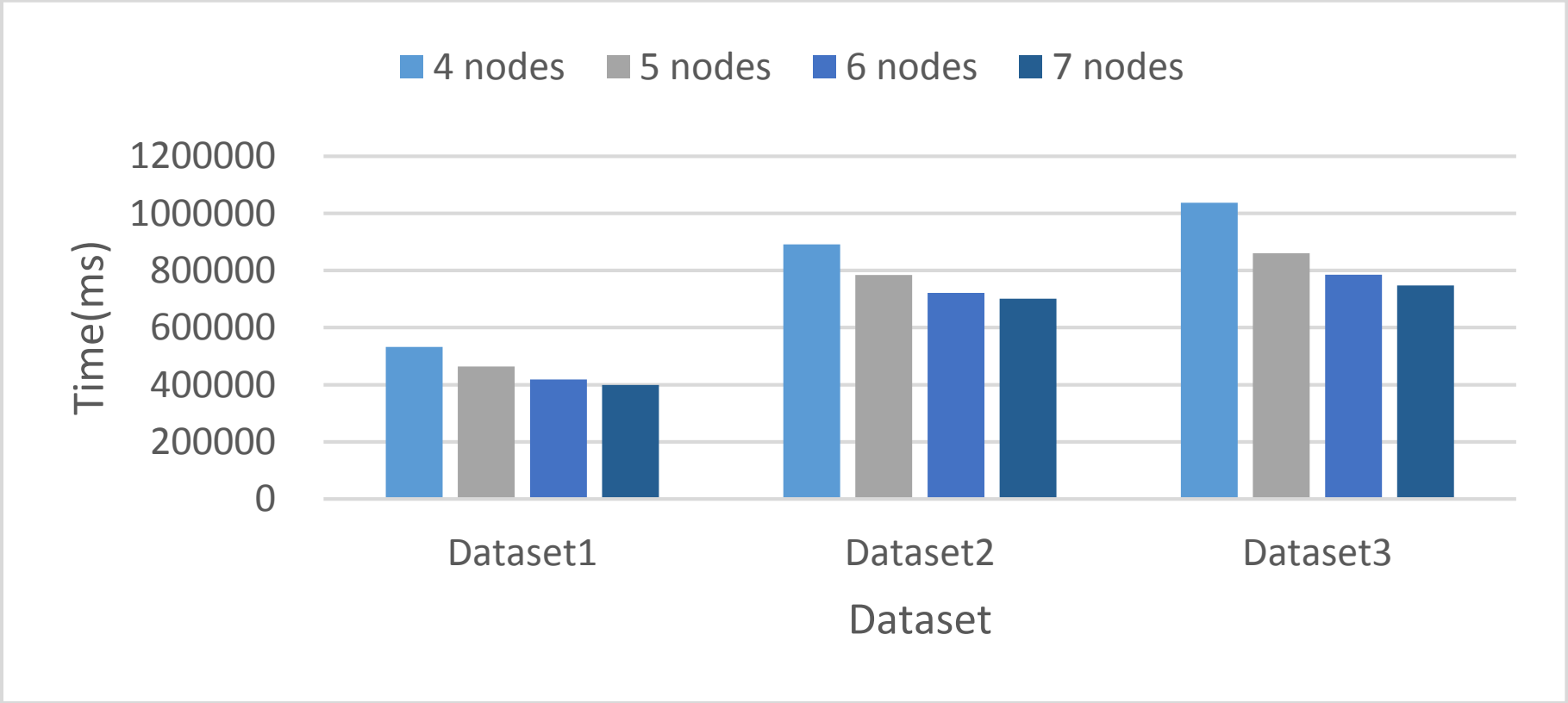

Fig. 3. Time consuming of proposed K-Medoids++ algorithm

The speedups for three different datasets are shown in Fig. 4. The performance of speedup of the proposed K-Medoids is very good according to the results. Specifically, the larger the size of the dataset is, the better the proposed K-Medoids spatial clustering algorithm performs. Consequently, the proposed parallel K-Medoids algorithm can process massive spatial datasets efficiently and scales well.

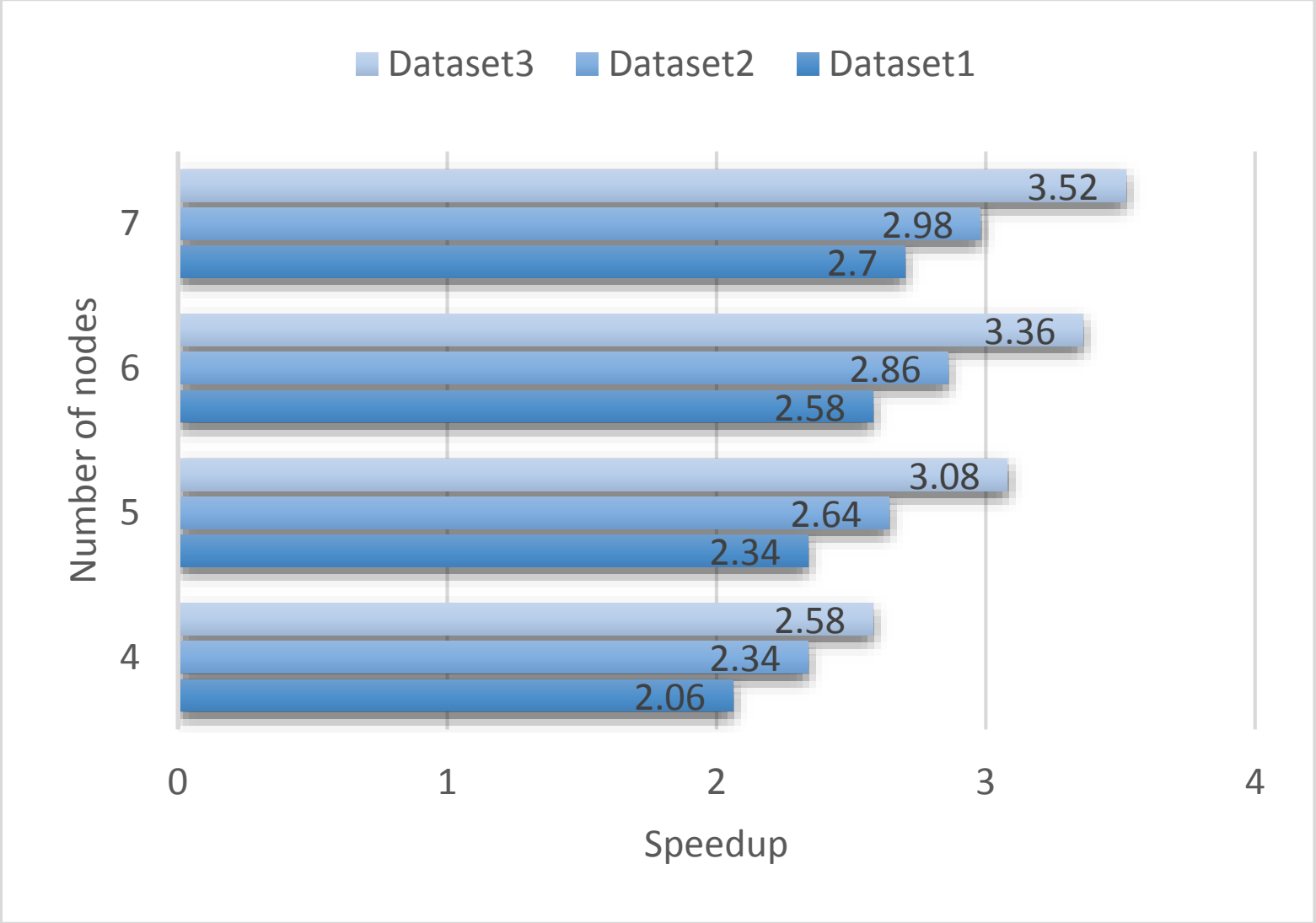

Fig. 4. Evaluation results

Finally, comparative experiments among different algorithms with different sizes of datasets were conducted. The classic clustering algorithms for comparison are traditional K-Medoids algorithm and CLARANS algorithm and the size of the three spatial data sets increases from 515MB to 1295 which are the same data sets listed below. The execution times of these comparative experiments and its visual illustration are shown in Fig. 5.

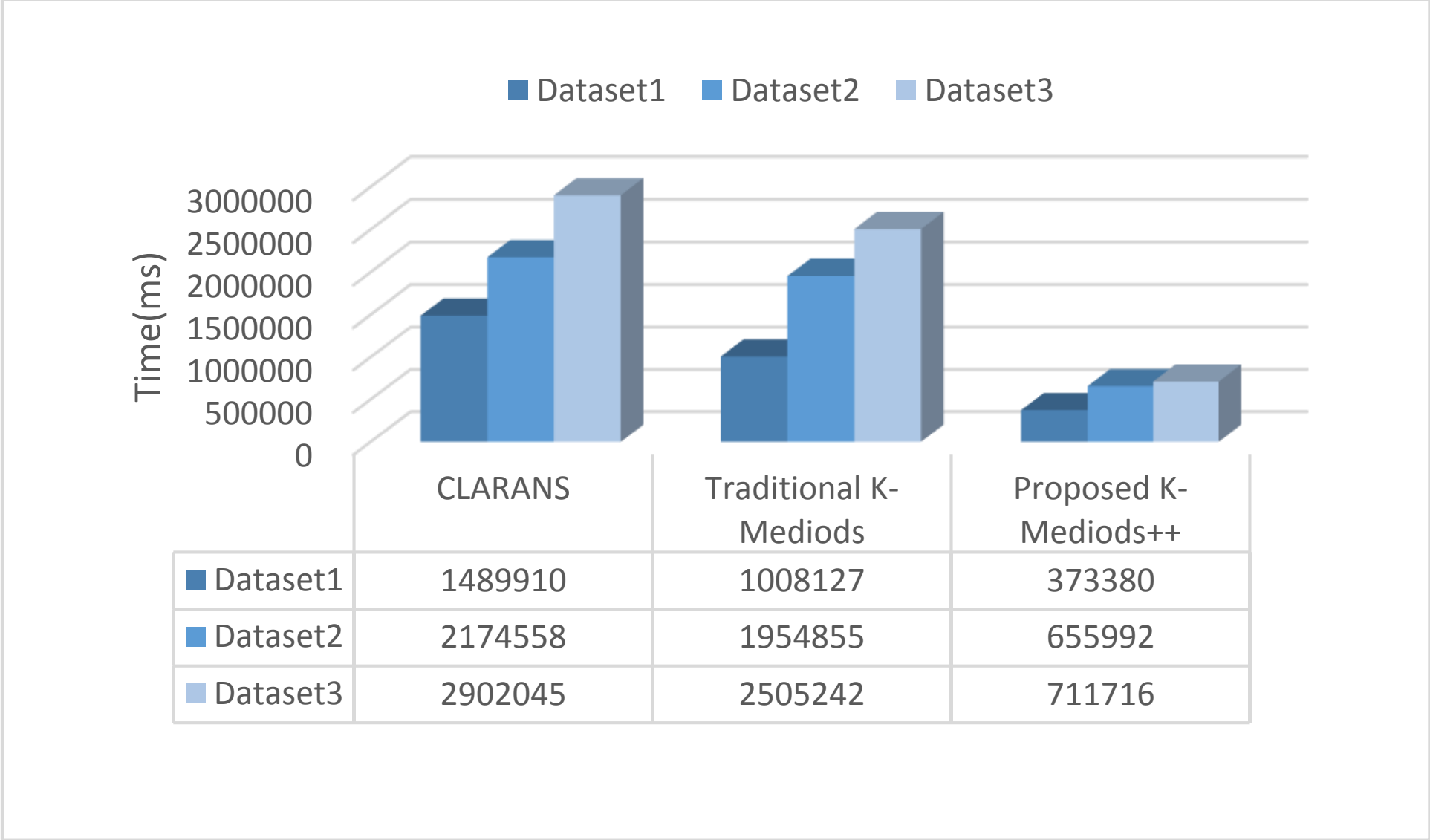

Fig. 5. Time Consuming of similar algorithm

## 5. Conclusions

Spatial clustering is of vital significance in geospatial data mining. The enlarging geospatial information of very large size makes spatial clustering a challenging task though many spatial clustering algorithms have been proposed in several decades. The techniques of parallel cloud computing are no doubt the suitable choices. In this paper, an efficient K-Medoids++ spatial clustering algorithm with the initial algorithm is proposed based on MapReduce framework which has been commonly embraced by academia and industry.

There are two main aspects of the improvement of the proposed K-Medoids++ spatial clustering algorithm. Firstly, the effective initial medoids searching algorithm is integrated into K-Medoids++ spatial clustering to find the suitable medoids as to decrease the number of iterations. Secondly, the K-Medoids++ spatial clustering algorithm is paralleled in MapReduce framework. 21 experiments are designed and the speedups of each group are used to evaluate the performances of the proposed algorithm. The experiment results show that the proposed K-Medoids++ spatial clustering algorithm can deal with massive spatial data effectively and scales well on commodity hardware.

## Funding

This paper is supported by the Science and Technology Project of State Grid Corporation of China (SGITG-KJ-JSKF[2015]0012).

## References

1. Han J., Kamber M. and Tung A., in: HJ Miller and J Han., Eds, Spatial clustering methods in data mining: A survey. In Geographic Data Mining and Knowledge Discovery, CRC Press, 2009: 12-30.
2. Zhang Q, Couloigner I. A new and efficient k-medoid algorithm for spatial clustering[M]//Computational Science and Its Applications–ICCSA 2005. Springer Berlin Heidelberg, 2005: 181-189.
3. MacQueen J. Some methods for classification and analysis of multivariate observations[C]//Proceedings of the fifth Berkeley symposium on mathematical statistics and probability. 1967, 1(14): 281-297.
4. Ball G H, Hall D J. ISODATA, "a novel method of data analysis and pattern classification," STANFORD RESEARCH INST MENLO PARK CA, 1965: 1-10.
5. Park H S, Jun C H. A simple and fast algorithm for K-medoids clustering[J]. Expert Systems with Applications, 2009, 36(2): 3336-3341.
6. Zhao W, Ma H, He Q. Parallel k-means clustering based on mapreduce[M]//Cloud Computing. Springer Berlin Heidelberg, 2009: 674-679.
7. Xiaoyang, "Estimating Language Models Using Hadoop and Hbase," University of Edinburgh, 2008: 3-22.
8. Arthur D, Vassilvitskii S. k-means++: The advantages of careful seeding[C]//Proceedings of the eighteenth annual ACM-SIAM symposium on Discrete algorithms. Society for Industrial and Applied Mathematics, 2007: 1027-1035.
9. S. Jianling, J. Qiang, "Scalable RDF Store Based on HBase and MapReduce," 3rd International Conference on Advanced Computer Theory and Engineering (ICACTE) , 2010, pp: 633-636.
10. Yuehu Liu, Bin Chen, Wenxi He, Yu Fang, "Massive image data management using HBase and MapReduce," Proceedings of the 21th International Conference on Geoinformatics, 2013, pp:1-5.
11. Kaufman L, Rousseeuw P J. Finding groups in data: an introduction to cluster analysis[M]. John Wiley & Sons, 2009: 2-40.
12. Deng min, Liu qiliang, Li guangqiang, Huang jianbo, Spatial clustering analysis and its application, Science press, 2011: 3-20.
13. Yue J, Mao S, Li M, et al. An efficient PAM spatial clustering algorithm based on MapReduce[C]//Geoinformatics (GeoInformatics), 2014 22nd International Conference on. IEEE, 2014: 1-6.
14. Yao K T, Lucas R F, Ward C E, et al. Data analysis for massively distributed simulations[C]//Interservice/Industry Training, Simulation, and Education Conference (I/ITSEC)(Got from Google Scholar). 2009: 2-32.